\documentclass[iop]{emulateapj}

\shorttitle{Science Interests with SKA}
\shortauthors{Kharb et al.}

\begin{document}
\title{From Nearby Low Luminosity AGN to High Redshift Radio Galaxies: Science Interests with SKA}
\author{P.~Kharb\altaffilmark{1,2}, D.~V.~Lal\altaffilmark{1}, V.~Singh\altaffilmark{3}, J.~Bagchi\altaffilmark{4},  C.~H.~Ishwara Chandra\altaffilmark{1}, A.~Hota\altaffilmark{5}, C.~Konar\altaffilmark{6}, Y.~Wadadekar\altaffilmark{2}, P.~Shastri\altaffilmark{2}, M.~Das\altaffilmark{2}, K. Baliyan\altaffilmark{3}, B.~B.~Nath\altaffilmark{7}, M. Pandey-Pommier\altaffilmark{8}}
\affil{$^1$National Centre for Radio Astrophysics - Tata Institute of Fundamental Research, Post Bag 3, Ganeshkhind, Pune 411007, India}
\affil{$^2$Indian Institute of Astrophysics, II Block, Koramangala, Bangalore 560034, India}
\affil{$^3$Astronomy \& Astrophysics Division, Physical Research Laboratory, Ahmedabad 380009, India}
\affil{$^4$Inter-University Centre for Astronomy and Astrophysics, Pune, India}
\affil{$^5$UM-DAE Centre for Excellence in Basic Sciences, Vidyanagari, Mumbai 400098, India}
\affil{$^6$Amity Institute of Applied Sciences, Amity University Uttar Pradesh, Sector 125, Noida 201303, India}
\affil{$^7$Raman Research Institute, C. V. Raman Avenue, Sadashivanagar, Bangalore 560080, India}
\affil{$^8$Univ Lyon, Univ Lyon1, Ens de Lyon, CNRS, Centre de Recherche Astrophysique de Lyon UMR5574,
9 av Charles André,  69230, Saint-Genis-Laval, France}

\begin{abstract}
We present detailed science cases that a large fraction of the Indian AGN community is interested in pursuing with the upcoming Square Kilometre Array (SKA). These interests range from understanding low luminosity active galactic nuclei in the nearby Universe to powerful radio galaxies at high redshifts. Important unresolved science questions in AGN physics are discussed. Ongoing low-frequency surveys with the SKA pathfinder telescope GMRT, are highlighted. 
\end{abstract}

\keywords{galaxies: active --- galaxies: Seyfert --- quasars: general --- BL Lacertae objects: general --- radio continuum: galaxies}

\section{INTRODUCTION}\label{intro}
Active galactic nuclei (AGN) are the centres of galaxies that emit copious amounts of radiation spanning the entire electromagnetic spectrum. It is now widely believed that AGN are accreting supermassive black holes (SMBH; masses $10^6-10^9$~M$_{\sun}$), where the enormous energy is the outcome of the release of gravitational potential energy \citep{Lynden69,HoKormendy00}. Somewhere in the interface between the SMBH and the accretion disk, bipolar jets or outflows are launched \citep[e.g.,][]{Rees82}. The details of the launch mechanism are still unclear, although magnetic fields are widely believed to be instrumental in the production and collimation of these outflows \citep{Blandford77,McKinney06,Tchekhovskoy11}. Strong and prominent emission lines in the optical-infrared spectrum, which are considered to be the hallmark of AGN, are produced in fast-moving gas clouds around the black hole-accretion disk system (the ``central engine"): different cloud speeds and electron densities have led to the demarcation into broad and narrow line regions (BLR, NLR). The BLR in some AGN are obscured from certain lines of sight by a dusty torus or a warped accretion disk, giving rise to the classification of type 1 (BLR and NLR visible) and type 2 (NLR visible but BLR obscured), and a unified scheme that attempts to link the two types on the basis of orientation \citep{Antonucci93,Netzer15}. Seyfert type 1s (with broad and narrow permitted emission lines in their spectra) and type 2s (with only narrow permitted and forbidden lines in their spectra) are expected to be the same phenomenon, differing only in orientation.

The development of radio interferometry in the 1960s led to the discovery of kiloparsec-scale jets in AGN. Large-scale jets that are observed in only about 15$-$20\% of AGN, emit at radio frequencies via the synchrotron process. Highly energetic electrons in these jets sometimes emit optical and X-ray synchrotron photons as well \citep{Sparks00,Worrall01,Harris06}. It was first pointed out by \citet{Fanaroff74} that kiloparsec-scale jets in radio galaxies exhibited primarily two radio morphologies: the Fanaroff-Riley (FR) type I galaxies had broad jets that flared into diffuse radio plumes $/$ lobes, while the FR type II galaxies had collimated jets that terminated in regions of high surface brightness called ``hot spots'' with the back-flowing plasma or the plasma left behind by the advancing jet, forming the radio lobes. The total radio power of the FRI and FRII sources also differed: the dividing line at 178 MHz was at $L_{178}\simeq2\times10^{25}$~W~Hz$^{-1}$, with the FRII sources being more radio powerful. The FR dichotomy is one of the major unresolved problems in jet astrophysics: it is not clear why the jets in FRIIs are powerful enough to produce hot spots, while they apparently lack power in FRIs. Differences in the mass or spin of the SMBH, accretion rate and/or mode, jet-medium interaction, host galaxy type and the galactic environments, are some of the explanations put forward to explain the FR dichotomy \citep{Prestage88,Baum95,Meier99,Hardcastle07}. 

The radio-loud unified scheme proposes that BL~Lac objects are the pole-on counterparts of FRI radio galaxies while the radio-loud quasars are the pole-on counterparts of FRII radio galaxies \citep{Urry95}. BL Lacs and quasars are collectively referred to as ``blazars''. Rapid variability at all wavelengths, high degrees of polarisation and superluminal jet motion, which are all the defining characteristics of blazars, are fully consistent with the suggestion of relativistic jets pointing close to our line of sight. Spectral energy distributions (SEDs) generated from multiwavelength studies of these blazars have revealed interesting trends: the synchrotron emission peaks at submm to IR ($\nu\sim10^{13}-10^{14}$~Hz) for the low frequency-peaked BL~Lacs (LBLs) and flat-spectrum radio quasars (FSRQs), while they peak at UV to X-rays ($\nu\sim10^{17}-10^{18}$~Hz) for the high frequency-peaked BL~Lacs (HBLs) \citep{Padovani95}. \citet{Sambruna96,Fossati98} and others have suggested that there are intrinsic differences in the physical parameters of these blazar sub-classes, with HBLs having higher magnetic fields/electron energies and smaller sizes than LBLs and FSRQs.

The vast majority of AGN however, do not exhibit powerful radio jets or outflows. Seyfert galaxies and LINERs\footnote{Low-Ionization Nuclear Emission-line Region} fall under this category. \citet{Kellermann89} classified AGN into the ``radio-loud" and ``radio-quiet" categories on the basis of their ratio ($R$) of radio flux density at 5~GHz to optical flux density in the $B-$band: $R$ was $\ll10$ for radio-quiet AGN. Differences in black hole masses, spins, accretion rates/modes, have been proposed to explain the radio-loud/radio-quiet divide \citep{Laor00,Tchekhovskoy10,Sikora07,Garofalo13,2010AJ....139.1089L}. According to the ``spin paradigm'', powerful radio jets originate near rapidly spinning accreting SMBH found in bulge-dominated systems, and are launched at relativistic speeds via the magnetohydrodynamic (MHD) Blandford-Znajek (BZ) mechanism \citep{Blandford77}. Alternatively, in the Blandford-Payne (BP) mechanism \citep{Blandford82}, jet power is extracted from the rotation of the accretion disk itself, via the magnetic field threading it, without invoking a rapidly spinning black hole. However, in both the processes the intensity and geometry of the magnetic field near the black hole horizon strongly influences the Poynting flux of the emergent jet \citep{Beckwith08}. This has given rise to the ``magnetic flux paradigm", which proposes that jet launching and collimation require strong magnetic flux anchored to an ion-supported torus of optically thin, geometrically thick, extremely hot gas with poor radiative efficiency \citep{Sikora13}. In general, a radiatively efficient fast accretion mode (or ``quasar'' mode) has typically been associated with a large fraction of ``radio-loud'' AGN, while a radiatively inefficient slow accretion mode (or ``radio'' mode) has typically been associated with ``radio-quiet'' AGN \citep{Croton06,Best12}. 

There is also evidence to suggest that galaxy mergers influence the radio-loudness of sources \citep{Heckman86,Deane14}. We know that galaxy mergers are an important phase of galaxy evolution \citep{Schweizer82,Mihos96,Cox06}. Mergers are especially important at high redshifts, where the galaxy density was higher compared to the present Universe. As all massive galaxies are expected to harbour SMBH in their centres \citep{Kormendy13}, galaxy mergers are expected to lead to a large fraction of binary black hole systems. Such systems have so far, however, been identified in only a handful of sources \citep[e.g.,][]{Deane14,Muller15}. Radio sources that exhibit double peaks in their emission line spectra are good candidates for binary AGN \citep{Liu10,Rosario11,Comerford13}. As are sources exhibiting highly distorted jets and lobes, called X-shaped sources \citep{Cheung07,Lal05b,Lal07}. Galaxy mergers are likely to be driving gas into the centres of merging galaxies, thereby providing fuel to be accreted on to the black holes. Yet not all dual AGN candidates exhibit one or two sets of bipolar kiloparsec-scale jets. The reasons for this absence are important but unclear. The search for dual AGN and a multi-wavelength study of dual AGN candidates can provide a different but unique perspective on the close interplay between the central SMBH and its host galaxy.

One way to detect dual AGN is through the technique of very long baseline interferometry (VLBI). For VLBI, unconnected radio telescopes located in separate parts of the world, work together to produce a single large radio telescope with milliarsecond resolution. This translates to parsec-scales at the distance of the AGN we aim to study. One of the most robust detections of a binary black hole system have been made by \citet{Rodriguez06} using multi-frequency VLBI. Moreover, multi-epoch VLBI observations are now determining jets speeds in blazars, radio galaxies, Seyferts and LINERs \citep[e.g.,][]{Ulvestad99,Lister09}. These studies have found that the jets in radio-loud AGN are typically faster than those in radio-quiet AGN. The reasons for the differences are not clear. Are jets being launched with different speeds by different mechanisms or are they getting slowed down by interacting with the surrounding medium\,? \citep[e.g.,][]{Bicknell86,Laing02}. The role played by jet-medium interaction in jet propagation is another important unsettled question in jet astrophysics. Bent radio jets in radio galaxies residing in galaxy clusters demonstrate the significance of jet-medium interaction \citep[e.g.,][]{ODea85,Lal13}, while VLBI observations suggest that jet differences occur close to the jet launching sites \citep[e.g.,][]{Lister09,Kharb10}.

High redshift radio galaxies (H$z$RGs) are known to reside in dense environments, and therefore, can be used as the tracers of (proto)clusters \citep{Venemans07,Galametz12}, as well as test cases for the study of jet-medium interaction. Since supermassive blackholes are essential ingredients of radio powerful AGN, the formation of SMBH at early epochs in the lifetime of the Universe, can also be probed with H$z$RGs. The host galaxies of H$z$RGs are among the most massive intensely star-forming galaxies and are believed to be progenitors of massive elliptical galaxies in the local Universe \citep{McLure04,Seymour07}. The radio luminosity function of H$z$RGs beyond redshift of 3 is poorly constrained. In fact, it is unclear whether there is a genuine dearth of H$z$RGs at $z>3$ or the observed deficiency is merely due to selection effects. H$z$RGs are also complementary to the emerging population of Lyman break galaxies at high redshifts, which are less massive by one to two orders of magnitude compared to the host galaxies of H$z$RGs. Therefore, the identification and study of H$z$RGs is important to understand the formation and evolution of galaxies at higher redshifts and in dense environments. 

Over the last two decades, supermassive black hole scaling relationships have become very well-established for both dormant and accreting SMBH. A clear corollary is strong coupling between the accretion processes that lead to a SMBH forming an AGN on the one hand, and star formation in their host galaxies on the other. When these accreting SMBH produce jets of plasma that reach hundreds of kpc, there  is clearly feedback into the intra-cluster medium as well. AGN are thus key players in our cosmic history, but  our physical understanding of the jet launch and feedback, has remained sketchy.  SKA will be able to sample like never before the parameter space of not only luminosity, redshift, jet collimation, quenching, time-domain behaviour and the black hole mass ladder, but also of the life-span of AGN activity. In the sections ahead, we discuss the work that has been carried out in different science areas by a large fraction of the AGN community in India. Sections are divided around the key unresolved science questions in AGN physics. We describe how SKA will help in resolving these science questions. {Details of upcoming SKA surveys and expected sensitivity levels are presented in Tables~\ref{tabskaA} and \ref{tabskaB}. 

In this article, we have adopted the spectral index convention, $S_\nu\propto\nu^{-\alpha}$, where  $S_\nu$ is the flux density at frequency $\nu$ and $\alpha$ is the spectral index.}

\begin{table}[t]
\small
\caption{SKA bands, including proposed changes in the low-frequency band definitions for SKA1-MID (Ref. ECP150027).}
\begin{tabular}{lllll}
\hline\hline
{}& {} & {MeerKAT} &  {SKA-Now} & {SKA-Final}\\ 
{} & {} & {$\nu$ (MHz)} &  {$\nu$ (MHz)}  & {$\nu$ (MHz)} \\ \hline
{SKA1-LOW} & {} & {\nodata}  & {50-350} & {\nodata} \\
{SKA1-MID} & {Band 1} & {580-1015} & {350-1050} & {475-875}\\
 {} & {Band 2} & {900-1670} & {950-1760} & {795-1470}\\
 {} & {Band 3} & {\nodata} & {1650-3050} & {1650-3050}\\ \hline
\end{tabular}
\vspace{0.35cm}
\label{tabskaA}
\end{table}

\section{JET FORMATION IN AGN}
Comparing and contrasting sources that possess jets with those that do not, is crucial for learning about jet formation. We discuss below how the ``radio-loud/radio-quiet'' divide can be, and is being tested on different fronts. First, sensitive radio observations are detecting kiloparsec-scale radio structures (KSRs) in Seyfert and LINER galaxies. These jets and lobes can extend from a kiloparsec to 10~kpc, or more. The host galaxies of these are typically lenticular or S0-type. Second, sensitive radio observations are detecting giant radio jets in spiral galaxies. Both these findings are challenging the previous well-accepted suggestions in the literature that large radio jets are only produced in elliptical galaxies, and not in spiral galaxies \citep{McLure99}. In the next two sections, we discuss these points individually. 

\subsection{KSRs in ``Radio-quiet'' Seyferts \& LINERs}
\citet{Gallimore06} have detected KSRs in $>44\%$ of Seyfert galaxies belonging to the volume-limited CfA$+12\mu$m sample when observed with the sensitive D-array configuration of the VLA at 5~GHz. \citet{Singh15a} found that $>43\%$ of Seyferts belonging to a  sample derived from the VLA FIRST\footnote{Faint Images of the Radio Sky at Twenty cm \citep{Becker95}} and NVSS\footnote{NRAO VLA Sky Survey \citep{Condon98}} surveys, possessed KSRs. Low frequency observations with the GMRT at 610~MHz are finding an even larger fraction of KSRs in Seyfert and LINER galaxies ($>50\%$, Kharb et al., in preparation). 

\begin{table}[b]
\small
\caption{Expected Sensitivity Levels of SKA bands.}
\begin{tabular}{lllll}
\hline\hline
 {}        & {} &  {Bandwidth}& {Field of View}& {{\it rms} noise}\\
 {}        & {} &  {(MHz)}  & {(deg$^2$)}& {($\mu$Jy~beam$^{-1}$)} \\ \hline
{SKA1-LOW} & {}           & {250}    & {39}       & {$\sim20^\dagger$} \\
{SKA1-MID} & {Band 1} & {700}    & {1.4}      & {$\sim$1.7}\\
 {}        & {Band 2}          & {810}    & {0.35}     & {$\sim$0.8}\\
 {}        & {Band 3}          & {1400}   & {0.12}     & {$\sim$0.7}\\ \hline
\end{tabular}
{Note: The listed sensitivity levels are achievable in one hour of integration time. $\dagger$ SKA1-LOW is limited by confusion noise. To circumvent this, one can use frequencies $\ge$200~MHz and eventually reach {\it rms} noise levels of $\sim$10~$\mu$Jy~beam$^{-1}$. }
\vspace{0.35cm}
\label{tabskaB}
\end{table}

While Seyfert galaxies have traditionally been categorised as ``radio-quiet'' AGN, \citet{Ho01} have demonstrated that when the optical nuclear luminosities are extracted through high resolution observations (e.g., from the {\it Hubble Space Telescope}) and the galactic bulge emission is properly accounted for by using specialized software like {\tt GALFIT}, then the majority of Seyfert galaxies shift into the ``radio-loud'' class. \citet{Kharb14a} have confirmed this trend in the Extended 12$\mu$m Seyfert sample. Moreover, they found a continuous distribution in the radio-loudness parameters of Seyfert galaxies and low-luminosity FRI radio galaxies. From their FIRST+NVSS VLA study, \citet{Singh15a} have found that $\sim$15\% of their Seyfert/LINER sample fall under the ``radio-loud" category, following the definition of \citet{Kellermann89}. 

We need to look closely at Seyferts with KSRs and low luminosity FRI radio galaxies with weak radio jets, which together constitute the low-luminosity ``intermediate'' AGN population. If accurate black hole masses can be estimated for all low-luminosity ``intermediate'' AGN, along with their core X-ray and radio luminosties, then they could be placed on the fundamental plane \citep[e.g.,][]{Merloni03}, to see where they lie. If their fundamental plane is offset from the rest of the classical objects, it may suggest a new mode of disk-jet coupling in these sources.

To produce statistically robust results, studies like these are most efficiently carried out using large galaxy samples drawn from large area radio surveys.
The SKA1-MID array has the potential to study weak KSRs in Seyfert, LINERs and low-luminosity FRI radio galaxies. The angular resolution that will be achieved ranges between $0.4''$ at 1.4~GHz (band 2) and $0.07''$ at 8.3~GHz (band 5). The three-tier survey at bands 1 and 2 (1 and 1.4~GHz) will also be ideal for carrying out this study \citep[see Table~1 in][]{Prandoni15}. It is expected that SKA will detect a much larger number of radio-``intermediate'' sources, that will fill the radio-loud/radio-quiet gap.  

\begin{figure}[t]
\centering
\includegraphics[width=8cm,trim=0 60 0 50]{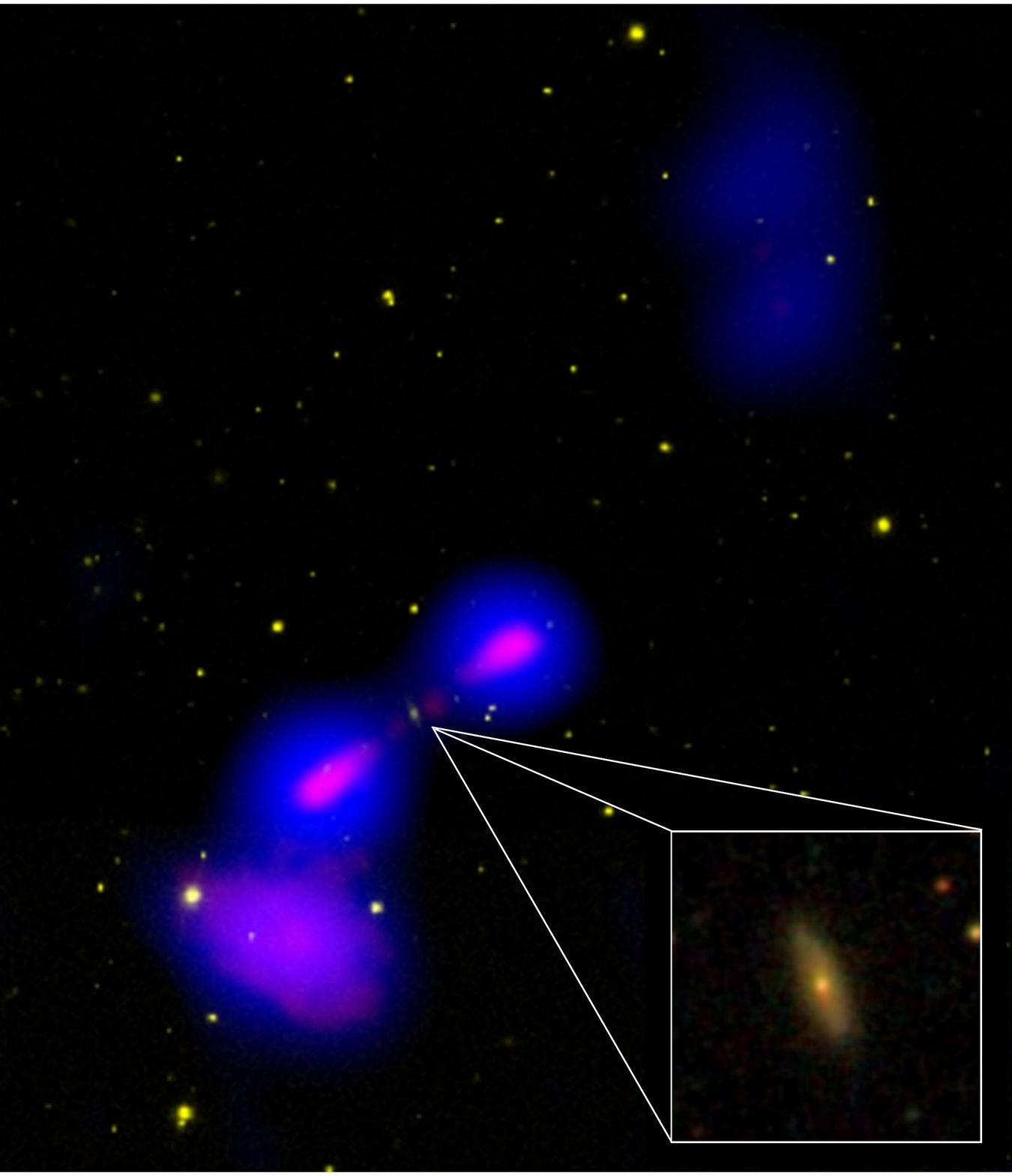}
\caption{The composite image of Speca \citep{Hota11}. The NVSS 1.4 GHz emission is in blue, while the GMRT 325~MHz emission is in red. The radio emission is superimposed on the optical image from SDSS. Inset shows the SDSS image of the host galaxy. }
\label{fig1}
\end{figure}

\subsection{Giant Radio Jets in Spiral Galaxies}\label{spiral}
With the advent of sensitive low radio frequency telescopes like the GMRT and LOFAR, several giant radio jets (extents $\gg100$~kpc) that are hosted in massive spiral galaxies, have been discovered \citep[e.g.,][]{Ledlow98,Hota11,Bagchi14,Mao15,Singh15b}. The standard paradigm to explain the elliptical hosts of large radio galaxies, invokes the process of two large spiral galaxies merging into an elliptical \citep[e.g.,][]{Schweizer86,Barnes92}. In this process, the central supermassive black holes of each spiral galaxy also merge and grow in mass. Therefore, the discovery of a few spiral galaxies hosting large or giant radio galaxies suggest alternate ways (other than mergers) of creating conducive physical conditions for the formation of radio galaxies. If the central black hole can grow big and spin fast, under whatever circumstances, it may be possible to create large radio jets.

\begin{figure}[b]
\centering
\includegraphics[width=7cm,angle=-90,trim=0 50 0 0]{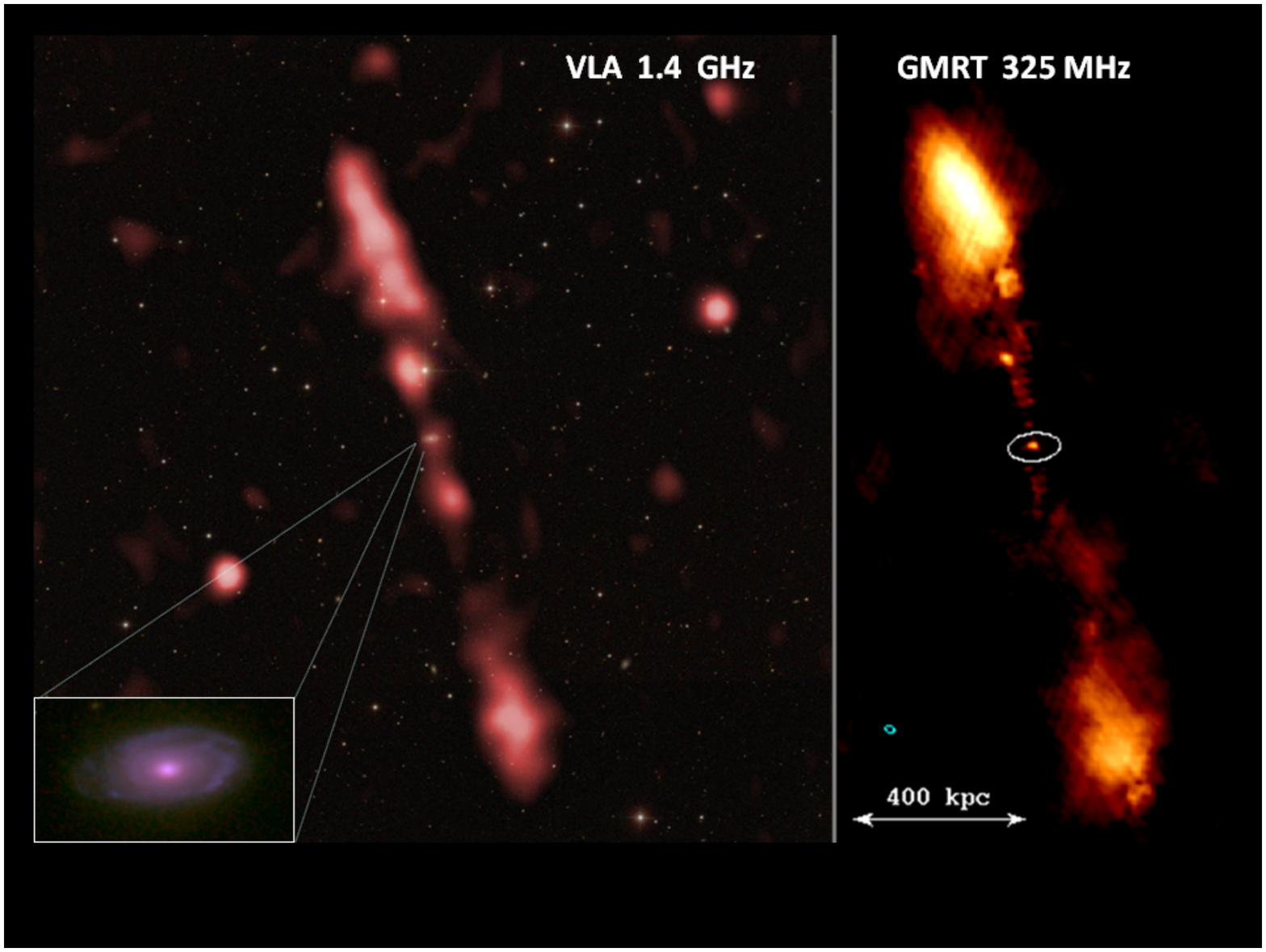}
\caption{The composite image of J2345$-$0449 \citep{Bagchi14}. The radio image from the NVSS at 1.4~GHz and the GMRT at 325~MHz are shown in red and orange. The inset shows the RGB-colour image of the host galaxy from CFHT. }
\label{fig2}
\end{figure}

The first large radio galaxy hosted in a disk galaxy (J0315$-$1906) was found by \citet{Ledlow98}. \citet{Hota11} reported the finding of the second such galaxy `Speca' (SDSS J140948.85$-$030232.5; see Figure~\ref{fig1}), using SDSS\footnote{Sloan Digital Sky Survey \citep{York00}}, NVSS and GMRT data. Recently, \citet{Bagchi14} reported the case of a spiral host in the source J2345$-$0449 (see Figure~\ref{fig2}). With the help of web-based citizen-scientists project (Radio Galaxy Zoo), \citet{Mao15} detected another such galaxy, J1649+2635, from an initial sample of 65,492 galaxies. Using similar automated archive cross-matching of SDSS, NVSS and FIRST data, \citet{Singh15b} have reported three new cases of spiral-host radio galaxies (viz., J0836+0532, J1159+5820, J1352+3126). Although J1649+2635 shows an optically blue grand-design spiral pattern, it is buried inside a huge old-stellar halo, unlike any spiral galaxy we know. Both J1352+3126 and J1159+5820 show clear signs of tidal interaction or past merger. So far, J0315$-$1906, Speca, J2345$-$0449 and J0836+0532 are the only four clear cases of large radio galaxies hosted by spirals.

Out of these four galaxies, two of them are giant and show episodic activity: J2345$-$0449 is a double-double radio galaxy of extent 1.6 Mpc, and Speca is likely a triple-double radio galaxy of extent 1.3 Mpc. J2345$-$0449 is a massive spiral galaxy (M$_\mathrm{dyn}\sim1.0\times10^{12}$~M$_\sun$), with a diameter of nearly 50 kpc, a fast rotation speed ($\sim$430~km~s$^{-1}$) and a high central velocity dispersion \citep[$\sigma\sim300$~km~s$^{-1}$;][]{Bagchi14}. Similarly, observations with the Subaru telescope have shown that Speca is large ($\sim$60 kpc) and has a rotation speed of $\sim$370~km~s$^{-1}$ \citep{Hota14}. Both these giant radio jets are therefore hosted by large and massive spiral galaxies. 

Such galaxies have likely grown in isolation without mergers but with co-planar accretion directly from cosmic-filaments or surrounding medium. They are expected to be more numerous at higher redshifts. None of these sources have so far shown an FRI-type radio morphology. This is likely to be due to the lack of sensitivity, especially for the high redshift sources. Once we have a larger number of such sources, the following questions can be answered: is there a difference in the bar fraction or metallicity in these host galaxies, or the presence of double nuclei~? Are there differences in the molecular gas fraction compared to other spiral galaxies ? Observing large samples of massive spirals with the SKA (see Section~\ref{grg} for more details), specifically at low radio frequencies, will increase the probability of finding several more spiral-host large radio galaxies, which have been missed in current surveys. 
Even before the operation of SKA1, the sample of such spiral galaxies will increase by two orders of magnitude due to optical spectroscopic surveys like DESI\footnote{Dark Energy Spectroscopic Instrument}.
High-resolution VLBI imaging and polarisation mapping of the inner radio jets near the core, in the disk/corona collimation region, would be very informative for understanding the jet launching process in these highly unusual AGN. 

Using the VLA FIRST survey, \citet{2011ApJS..194...31P} and \citet{2001A&A...371..445M} have listed three dozen FRII or FRI/II sources with angular sizes between brightest regions $\gtrsim$3$^\prime$ and a flux density limit, from a 3000 deg$^2$ region. Extrapolating from these detections, and using Figure~1 from \citet{2015aska.confE.101J}, we expect the SKA  surveys to detect about a few times 10$^4$ giant radio galaxies (see also Kale et al. 2016 in these proceedings). If we make a conservative estimate that $\sim$10\% of these are in spiral hosts, we will observe a thousand new giant radio galaxies hosted by spirals. An all-sky radio continuum survey using SKA1 will provide statistically large samples so that we can finally study them as a distinct class of AGN.

\section{JET MORPHOLOGY: CLUES TO FORMATION AND PROPAGATION}
Jets in AGN span parsec to mega-parsec scales, and have speeds ranging from a few thousand km~s$^{-1}$ to a fraction of the speed of light. Their morphologies are diverse: they can be bent, S- or Z-shaped, flaring or highly collimated. They can possess terminal hotspots or be plume-like with no clear jet termination. These morphologies can indicate differences in the jet launch speeds and/or jet-medium interaction. Attempts to unify this large variety of jet structures have been made. We discuss ahead the successes and failures of these attempts.

\begin{figure}[t]
\centering
\includegraphics[width=8.4cm]{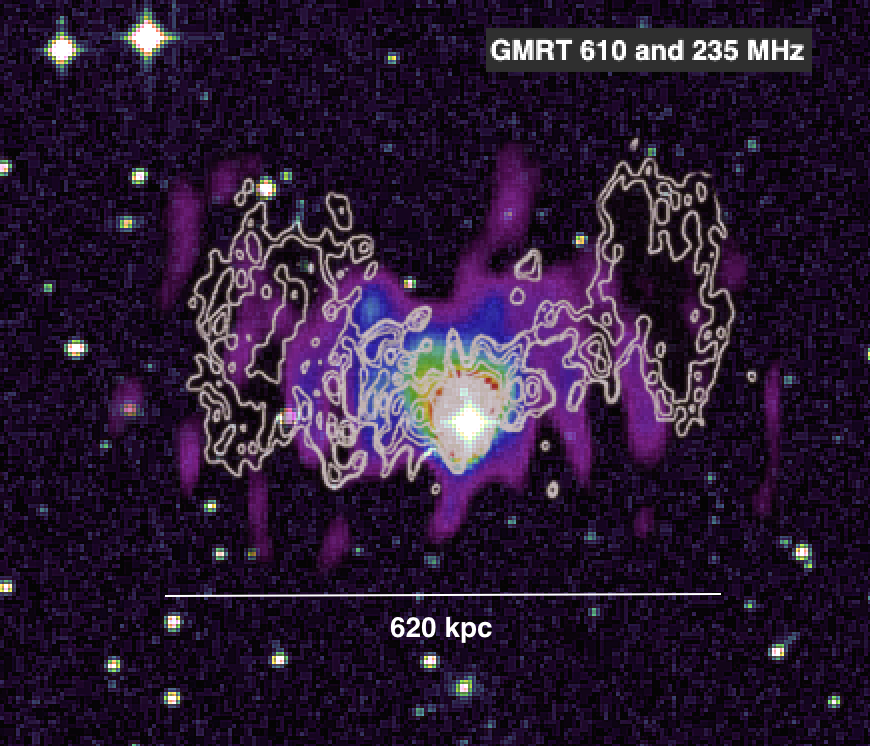}
\caption{Sensitive GMRT observations at 235 and 610 MHz (in colour and contours) reveal a $\sim$600 kpc-scale radio jet in the erstwhile core-dominated BL~Lac object PKS~2155$-$304 \citep{Pandey16}.}
\label{fig3}
\end{figure}

\subsection{The FR Divide \& Radio-loud Unification}
\begin{figure}[t]
\centering
\includegraphics[width=9cm,trim=70 210 10 220]{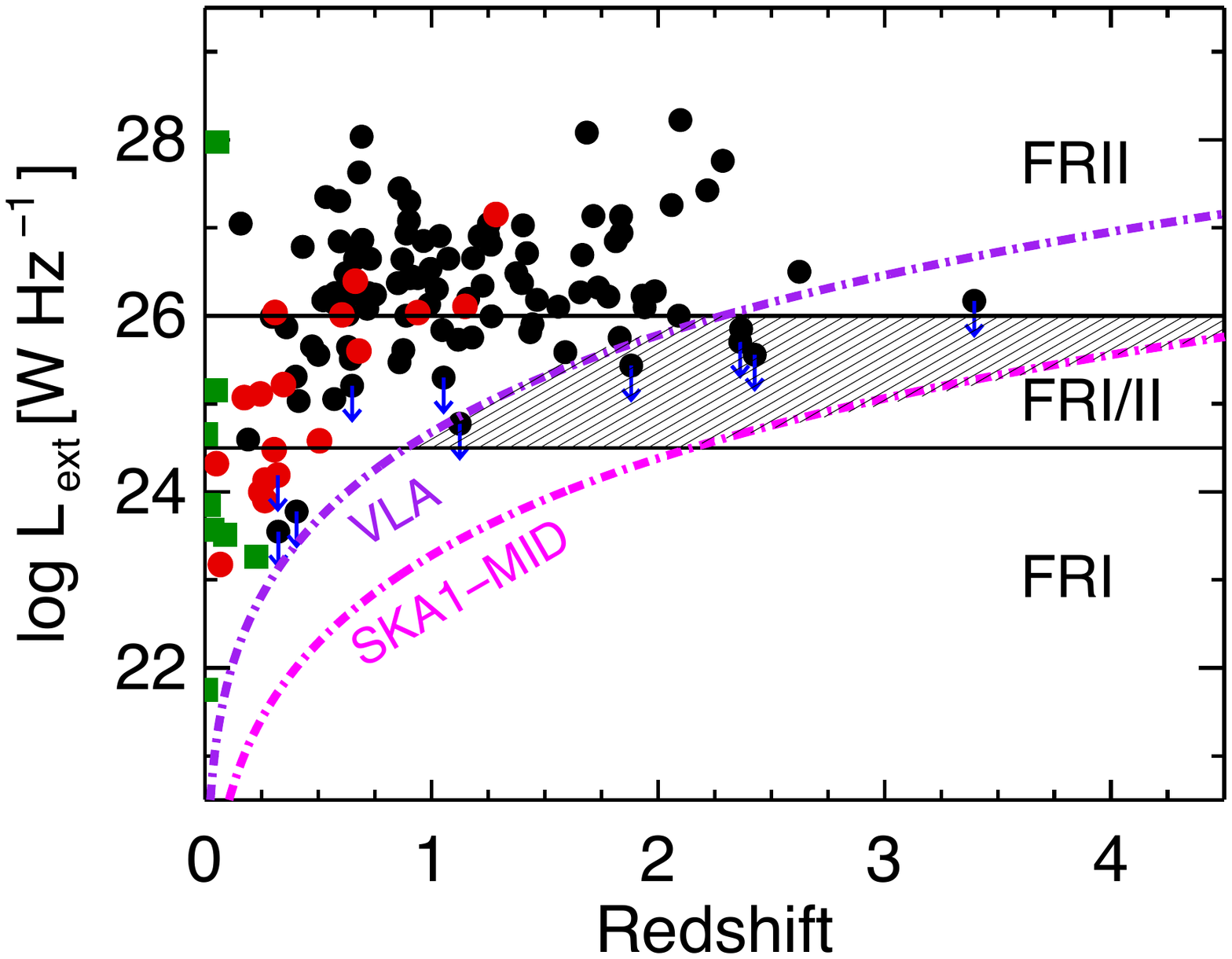}
\caption{1.4~GHz extended luminosity versus redshift for the MOJAVE sample \citep{Kharb10}. Black and red circles denote quasars and BL Lac objects, respectively, while green squares denote radio galaxies. Core-only sources are represented as upper limits with downward arrows. The solid lines indicate the FRI$-$FRII divide (extrapolated from 178 MHz to 1.4 GHz assuming a spectral index, $\alpha$=0.8), following \citet{LedlowOwen96} and \citet{Landt06}. The purple line denotes the sensitivity limit for the historical VLA which was used to carry out the MOJAVE study, while the magenta line denotes the sensitivity limit for the upcoming SKA1-MID array \citep[see Table~1;][]{Dewdney13}. SKA will be able to detect nearly twice as many ``intermediate'' or ``hybrid FRI/II'' sources, compared to previous studies.}
\vspace{0.2cm}
\label{fig4}
\end{figure}

The simple radio-loud unified scheme linking BL~Lac objects to FRI and quasars to FRII sources, has been challenged by \citet{Kharb10} on the basis of their kiloparsec-scale radio study of the MOJAVE\footnote{Monitoring of Jets in AGN with VLBA Experiments. http://www.physics.purdue.edu/MOJAVE} blazar sample. Kharb et al. examined the MOJAVE sample of 135 blazars \citep{Lister09}, using high-resolution 1.4~GHz data from the VLA, and found that a substantial fraction ($\approx20\%$) of MOJAVE quasars and BL Lacs had total radio powers that were ``intermediate'' between FRIs and FRIIs (see Figure~\ref{fig4}). Many BL Lac objects had lobe luminosities ($\approx30\%$) and hot spots ($\approx60\%$) like quasars. In addition, they found a strong correlation between the kiloparsec-scale lobe luminosities and parsec-scale jet speeds: the large-scale jet and lobe knew about the small-scale jet as it was launched. It therefore appeared that the fate of the AGN was decided at its birth!

These results have important implications for the inner workings of AGN. Therefore, they need to be re-examined with higher sensitivity radio data, as will become available with SKA. For instance, Figure~\ref{fig3} shows an example of an FRI BL~Lac object which was earlier known to host a highly variable compact radio core, but reveals extensive diffuse emission in a jet extending to 620 kpc, in a deep GMRT study at 235 and 610~MHz \citep{Pandey16}. 
{Could some BL Lac objects shift into the FRII luminosity category (contrary to the expectations of the Unified Scheme) with the detection of diffuse lobe emission ?}
As is clear from Figure~\ref{fig4}, the much lower sensitivity limit of SKA1-MID at GHz frequencies, is likely to detect nearly twice as many ``intermediate'' or hybrid FRI/II sources \citep[e.g.,][]{2010ApJ...722.1735L,Stanley15}, than previous studies carried out with the VLA. Our (historical) VLA observations at 1.4~GHz could not detect extended radio emission below $\sim$10$-$50~$\mu$Jy~beam$^{-1}$, but SKA1-MID will be able to detect emission that is as low at 0.7~$\mu$Jy~beam$^{-1}$ \citep[Table~1;][]{Dewdney13}, thereby increasing the overall sensitivity by a factor of 10 to 70. 

These sensitive data will become available for a much larger population of radio sources including the so-called ``core-only'' sources \citep[$\approx7\%$ in the MOJAVE complete sample;][see upper limits in Figure~\ref{fig4}]{Kharb10}. There are suggestions that these ``core-only'' blazars are another beast altogether \citep[e.g.,][]{Punsly15}. SKA will find out whether this is a new class of sources or sources with very faint extended emission beyond the reach of present day radio telescopes. Perhaps they exhibit episodic activity and are currently switched off: the faint emission from the previous episodes is not detectable. 

The SKA1-LOW and SKA1-survey\footnote{We note that SKA1-survey in Australia has been deferred in the SKA1 re-baselining of March 2015; see \url{https://www.skatelescope.org/wp-content/uploads/2014/03/SKA-RBS-outcome.pdf}} are ideal for detecting the full extent of the diffuse lobe emission in radio galaxies and blazars. SKA1-LOW is expected to operate between $0.05-0.35$ GHz, while the SKA1-survey will operate between $0.65-1.67$~GHz \citep[Table~1;][]{Dewdney13}. The resolution of SKA1-LOW would typically be around 11 arcsec, while the SKA1-survey will be around 0.9 arcsec. The three-tier survey at bands 1 and 2 (1.0 and 1.4~GHz) will be ideal for carrying out this project \citep{Prandoni15}. 

\subsection{Probing Blazar Nuclei with SED Modelling}
A large number of blazars (radio-loud quasars and BL~Lac objects) are being detected in the {\it Fermi} gamma-ray survey of the sky. For many sources, it is difficult to find their association in various energy regimes and their identification (e.g., LBL, HBL) due to lack of extensive data. Multi-frequency observations at all frequencies along with optical and radio polarization measurements are required: while one can use polarization information to identify and broadly classify blazar classes, continuum fluxes at all possible energy regimes are required to generate the SEDs. For instance, the blazar candidate CGRaBS~J0211+1051 was detected by {\it Fermi} in a flaring mode. \citet{Chandra12} carried out optical polarization observations of this source with the Mount Abu Infrared Observatory, India, in 2011, and detected high and variable degree of polarization (9$-$20\%). They therefore proposed the source to belong to the LBL class. To confirm this, a multi-frequency study was carried out for this source using data from the MOJAVE, {\it Planck}, {\it WISE}, {\it Swift}, Mount Abu Observatory, {\it Fermi} and {\it VERITAS}. The SED showed that low energy (synchrotron) peak fell at  $10^{14}$~Hz, confirming it to be a low-energy peaked blazar. The light-curves showed variations in the high energy gamma-rays to be correlated with X-ray, UV and optical variations, indicating their co-spatial origin \citep{Chandra14}. Multi-band continuum data from SKA will greatly benefit generation of SEDs for blazars while VLBI data will be useful for localising the emission regions in jets. {Classifying a greater number of blazars into various sub-categories can help us isolate the essential emission or physical mechanisms that are in play in these sources.} 

\subsection{Estimating Radio Lobe Energetics}
X-ray observations of radio galaxy lobes have shown that the electrons responsible for the synchrotron radiation in radio wavelengths can also scatter the cosmic microwave background (CMB) radiation photons to X-ray wavelengths. \citet{Tamhane15} discovered inverse Compton scattering by electrons in the lobes of a high redshift giant radio galaxy. CMB up-scattering is easier to detect at higher redshifts. This up-scattering requires that the Lorentz factors of the electrons be of order $\sim 10^4$, which implies that the regions in the lobe which would be X-ray bright would also be bright at low radio frequencies (few hundred MHz), for typical values of magnetic fields \citep{Nath10}. However, the surface brightness of these regions is likely to be low. Observations with SKA would be useful in determining the physical processes at work in the radio lobes (back flow, cooling of electrons due to adiabatic expansion of lobe, through radiation etc.).

\section{JET KINEMATICS AND MAGNETIC FIELD STRUCTURE}
Differences in jet launching speeds and magnetic field structures are widely believed to be important in explaining the wide variety of jets observed in AGN. Multi-epoch VLBI observations are essential to measure jets speeds on parsec and sub-parsec scales. We discuss multi-epoch-VLBI below in Section~\ref{sec-speed}. The feasibility of carrying out VLBI and polarization-sensitive VLBI with SKA (or SKA-VLBI) has been discussed by \citet{Paragi15}. Polarization-sensitive VLBI (VLBP) observations can estimate the projected magnetic field structures in parsec-scale jets \citep{Gabuzda94}. Polarisation measurements in kiloparsec-scale jets and lobes can reveal the effects of jet-medium interaction on inter-galactic scales \citep{Liu91,Laing96,Kharb12}.

\subsection{Jet Speeds from Multi-epoch VLBI}\label{sec-speed}
Using VLBI images of a rigorously selected sample of Seyfert galaxies, \citet{Lal04} have shown that the radio properties of the compact parsec-scale features are consistent with the unified scheme for Seyfert galaxies (see Section~\ref{intro}), with no significant evidence for relativistic beaming in their jets \citep[see also][]{Lal11}. Some parsec-scale features in these sources could be termination points of radio jets. The high-resolution, parsec-scale study of low-luminosity AGN is challenging because these sources have very low radio flux densities, typically less than a few milli-Jansky. Therefore the correlated flux densities of only a handful of sources meet the sensitivity thresholds currently offered by global VLBI, which can currently only study AGN with flux denisties of least a few milli-Jansky. SKA-VLBI will be able to decrease this flux density limit by several orders of magnitude. This will allow us to study jets and jet speeds in low luminosity AGN, just as they are launched from the black hole - accretion disk systems.

\subsection{Magnetic Fields on Parsec-scales}
VLBP observations of FRI radio galaxies at 8~GHz with a global VLBI array by \citet{Kharb05} were among the first observations carried out on this relatively weak class of sources. Almost all previous such observations had been carried out on the more radio powerful FRII radio galaxies and blazars. \citet{Kharb05} found a 100\% detection rate of jet polarization in the four FRI radio galaxies they observed and a suggestion of a ``spine-sheath'' magnetic field structure. Furthermore, the detection of significant core polarization was consistent with FRI radio galaxies lacking the putative dusty torus, the inner ionized region of which should have depolarized the core emission. The global VLBI array used in this study included the 100 m Effelsberg antenna: without the inclusion of this large single dish telescope, the VLBI array would not have been as  sensitive and the success rate of detection would not have been 100\%.
A multi-frequency (5, 8, 15 GHz) VLBP study of these sources revealed a gradient in rotation measure (RM) across the jet of the FRI galaxy, 3C\,78. Such a gradient suggests the presence of a helical magnetic field \citep{Blandford93,Kharb09,Hovatta12,Gabuzda14}. Similar multi-frequency VLBP observations of a complete sample of UGC FRI radio galaxies, revealed sheath-like magnetic field structures in several sources, which are also consistent with helical magnetic fields \citep{Kharb12} (see also \citet{Laing93,Reichstein10}). 

SKA-VLBI with SKA1-MID and/or SKA1-SUR, observing together in the $1-15$~GHz frequency range with the current VLBI arrays and future large aperture telescope like FAST \citep{Nan11}, is expected to outperform current global VLBI arrays including even the most sensitive current telescopes \citep{Paragi15}. Faraday rotation measurements and the search for RM gradients in both the approaching and receding jets of relatively-plane-of-sky radio galaxies of the FRI and FRII types, need to be carried out, in order to probe the complex magnetic field structures in jets, without relativistic effects that can dominate in the low inclination blazars. Since radio galaxies are much weaker than blazars in total radio emission, this study can only be carried out for a large enough number of sources with the future SKA-VLBI.

\subsection{Magnetic Fields on Kiloparsec-scales}
Multi-frequency polarization-sensitive observations with the VLA of a sample of 13 FRII radio galaxies, revealed interesting correlations on kiloparsec-scales \citep{Kharb08}. A strong correlation between lobe depolarization and lobe spectral index (``Liu-Pooley effect'') was observed: radio lobes with a flatter spectrum exhibited lower depolarization. The lobe depolarization difference was correlated with the arm-length ratio: the shorter lobe in the source was more depolarized \citep[see also][]{Pedelty89,Laing96}. This strongly suggested that lobe depolarization depends significantly on environmental asymmetries in radio galaxies. Most interestingly, \citet{Gabuzda15} have recently found suggestions for an ordered toroidal magnetic field component in the AGN 5C\,4.114 on kiloparsec scales, through RM observations. Arcsecond-scale polarimetric observations can therefore provide important clues about the kiloparsec-scale environments of radio galaxies and blazars, as well as infer the presence of large-scale magnetic fields in their jets and outflows. {SKA will detect fainter polarization signals on smaller spatial scales in the cores, jets, lobes and hotspots of radio galaxies, than is feasible with the current radio telescopes. This will be crucial in detecting signatures of ordered magnetic fields on a range of spatial scales in radio galaxies, or examining the intervening medium between the galaxies and us in much finer detail, with RM data.}

\section{AGN LIFETIMES AND DUTY-CYCLE}
The lifetimes of radio sources have been inferred from several arguments. Spectral ageing analysis has been a powerful tool to infer radio source ages. Giant radio galaxies (GRGs) push the limits of the lifetimes for radio sources. It is also clear that AGN activity is episodic in nature \citep[e.g.,][]{Davies90,Saikia09}. Attempts have been made to infer the AGN duty-cycle \citep[e.g.,][]{Greene07}. These are essential to understand why only a fraction of all galaxies in the Universe have nuclei that are ``active'', even though all massive galaxies host supermassive black holes. 

{There have been attempts to investigate the lifetime and duty-cycle of AGN activity by studying radio galaxies exhibiting two pairs of lobes that are formed during two different phases of AGN activity \citep[e.g.][]{Jamrozy09,Konar13}. It is important to note that despite the availability of thousands of radio galaxies in various radio surveys, only a few dozen are confirmed `double-double radio galaxies' \citep[DDRGs;][]{Schoenmakers00,Nandi12}. The paucity of DDRGs can be understood as a lack of radio data with optimum sensitivity and resolution to detect radio structures of different spatial scales and surface brightness. Using deep 610 MHz GMRT observations, \cite{Singh16} recently reported the discovery of a `triple-double radio galaxy' (TDRG) J1216+0709, which exhibits three pairs of lobes from three different AGN activity episodes. The different structures seen in the GMRT image are undetected in all the previous radio surveys (e.g., FIRST, NVSS, VLSS, TGSS) due to the lack of optimum sensitivity and resolution. Therefore, more sensitive multi-frequency, multi-resolution observations from SKA are expected to detect radio structures from episodic AGN activity, that have so far remained undetected.}

\subsection{Giant Radio Galaxies}\label{grg}
GRGs can be used as pointers of black hole physics and `barometers' of the intergalactic medium in the cosmic-web \citep[e.g.,][]{Nath95,Malarecki13}. GRGs, whose lobes span $\sim$1 Mpc or more are among the largest, most luminous objects in the Universe \citep{Ishwara99}. Due to sensitivity limitations of the present day radio telescopes, most of the GRGs found so far are in the nearby universe ($z\lesssim0.7$). Recently, Sebastian et al. (2015, in preparation) have discovered a 2.2 Mpc giant radio source at a redshift of 0.56 in the field of LBDS-Lynx\footnote{Leiden-Berkeley Deep Survey - Lynx}. This appeared as a faint elongated source in the deep 150 MHz images with the GMRT, implying that the deeper radio surveys at low frequencies such as SKA-LOW have the potential to discover many giant radio sources at redshift $>0.5$. 

It is unclear if the large sizes of GRGs reflect the high efficiency of radio jets produced from the central engine, or they grow to enormous sizes due to their favourable location within a low density ambient medium. Approximately half of the baryons in the present day Universe are still unaccounted for (`missing'), in the sense that these baryons are believed to reside in the large galaxy filaments, in the form of warm-hot intergalactic medium (WHIM), as part of the cosmic-web structure of the universe \citep{Dave01,Werner08}. The large extents of GRGs provide an excellent opportunity to use them as barometers for probing the physical properties of this WHIM gas (its temperature, pressure and magnetic field). For this purpose a sensitive search using SKA and LOFAR for Mpc scale radio sources in the vicinity of galaxy filaments and perhaps inside the voids, surrounded by sheets of galaxies would be extremely valuable.
The discovery of large numbers of giant radio galaxies in the distant universe will be greatly facilitated by the concurrent operation of SKA1 and LSST\footnote{The Large Synoptic Survey Telescope}. Accurate photometric redshifts for millions of galaxies provided by LSST will be critical in this effort. 

In Section~\ref{spiral}, we had estimated the number of GRGs likely to be detected with SKA surveys by extrapolating from the detections seen in VLA FIRST survey images \citep{2011ApJS..194...31P,2001A&A...371..445M} and Figure~1 from \citet{2015aska.confE.101J}. To recapitulate, an all-sky radio continuum survey using SKA1 will provide statistically large samples of a few times 10$^4$ GRGs.

\subsection{Double-double Radio Galaxies}
DDRGs exhibit two episodes of jet activity, where a new pair of jets plows through the cocoon material dumped by previous jet activity. The duration of the active phase is a few to a few 100 Myr, and the duration of the quiescent phase is a few $10^5$ to a few $10^8$ yr \citep{Konar13a,Marecki09}. As the outer relic lobes are expected to be steep spectral plasma, a few new DDRGs have been discovered using the low radio frequency observations with the GMRT: detailed spectral ageing analyses have been performed after incorporating complementary high-frequency data from the VLA \citep[e.g.,][]{Saikia06,Konar06,Nandi14}. \citet{Tamhane15} have discovered a new Mpc-sized radio galaxy at $z$=1.32 in the XMM-LSS field. {This is a relic radio galaxy exhibiting co-spatial radio and X-ray lobe emission, but no core, jets or hotspots.}

An important question in DDRGs is whether the jet power remains similar in two consecutive episodes. \citet{Konar13b} found strong evidence that the jet power, at least in a fraction of DDRGs, remains similar in two consecutive episodes: the injection spectral indices of inner and outer lobes were similar and an $\alpha^{in}_{inj}-\alpha^{out}_{inj}$ correlation between the injection spectral indices of inner and outer double, was detected. \citet{Konar13b} however, were studying only a few sources. With SKA, we expect to discover hundreds of DDRGs, so that one can verify the $\alpha^{in}_{inj}-\alpha^{out}_{inj}$ correlation and check for its universality. \citet{Konar13b} also found that the smallest duration of quiescent phase between two episodes (so far) was $\sim10^5$ yr. A statistical study to find the distribution of quiescent phases of DDRGs is important to constrain the accretion physics of the central engine. With the expected discovery of many new DDRGs with SKA, we hope to study such a distribution.

\begin{figure}
\centering
\includegraphics[width=17cm]{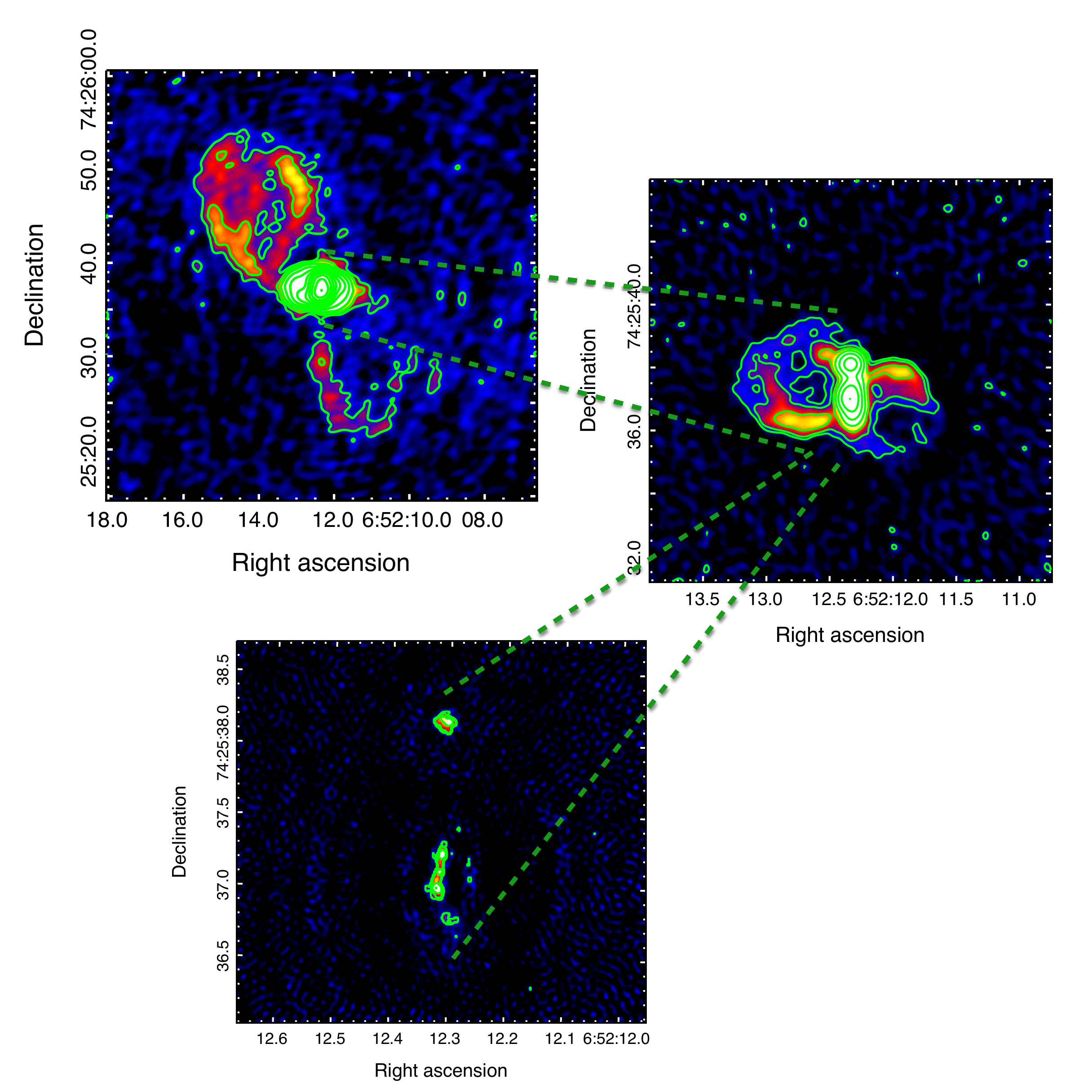}
\caption{The Seyfert galaxy Markarian 6 exhibits three sets of radio lobes/jets which are nearly perpendicular to each other \citep{Kharb06}. The detection of these structures required sensitive observations with multiple arrays and frequencies with the VLA and MERLIN. Steep-spectrum radio lobes from previous episodes of the AGN activity require sensitive low frequency observations, as will become available with SKA. Many such complex radio morphology sources are likely to be detected then.}
\label{fig5}
\end{figure}

Another important goal is to find the oldest plasma around the outer double of the known DDRGs. So far, the oldest known plasma around a DDRG is 200~Myr in the source 4C\,29.30 \citep{Jamrozy07}. {Since synchrotron ageing is a slow process at low electron energies, responsible for $\sim$100~MHz radio emission, SKA will be able to detect radio relics that are as old as 100 million years.} SKA is also likely to discover more triple-double radio galaxies. So far, there have only been {three cases: B0925+420 \citep{Brocksopp07}, Speca \citep{Hota11} and J1216+0709 \citep{Singh16}}. In the Seyfert galaxy Mrk\,6, three  episodes are seen in three different orientations \citep[][see Figure~\ref{fig5}]{Kharb06}. In NGC\,5813, three pairs of cavities in X-ray emission have been found: these were later found to contain old relativistic plasma in GMRT observations \citep{Randall11}. From the top left panel of Figure~1 of \citet{Prandoni15}, we see that the 5$\sigma$ SKA {\it rms} noise at $\sim$100 MHz is $\sim$0.2 mJy~beam$^{-1}$. This is much lower than most of the existing surveys, as well as {\it rms} noise of GMRT images (1$\sigma$ $\sim$a few mJy~beam$^{-1}$) at 150~MHz after 8 hours of integration. 
The higher resolution of SKA will probe the jet-lobe structure well. This will be useful to discover new aspects of jet physics, as well as to verify the work discussed here.

\citet{2011ApJS..194...31P} and \citet{2001A&A...371..445M}, have listed 242 systems including triple-double systems and over $\sim$600 more candidates, in the VLA FIRST survey. \citet{2009MNRAS.395..269S} have listed 374 sources from an 18$^\prime$ square ELAIS-N1 field using the GMRT at 325~MHz up to a median {\it rms} noise of $\sim$40~$\mu$Jy and an angular resolution of $\sim$8$^{\prime\prime}$. SKA1 surveys will be more sensitive than the GMRT ELAIS-N1. Extrapolating from  these detections, and using Figure~1 from \citet{2015aska.confE.101J}, we expect that the SKA surveys will detect $\sim$10$^7$ DDRGs. 

\subsection{Episodic Activity in Low-luminosity AGN}
Observations of the Seyfert galaxies NGC\,4235 and NGC\,4594 (the Sombrero galaxy), with the GMRT at 325 and 610 MHz, have revealed signatures of episodic AGN activity in them \citep{Kharb16}. Both the 610~MHz total intensity and the $325-610$~MHz spectral index images suggest the presence of a ``relic'' radio lobe in NGC\,4235, suggestive of episodic activity in this galaxy. Based on a simple spectral ageing analysis, the relic outer lobe appears to be at least two times older than the present lobe. This implies that the AGN in NGC\,4235 was switched ``off'' for the same time that it has been ``on'' for the current episode. A $\sim3$~kpc linear, steep-spectrum ``spur-like'' feature is observed nearly perpendicular to the double-lobed structure in NGC\,4594. Since the VLBI jet in NGC\,4594 is perpendicular to this linear feature, the AGN does not seem to be currently fuelling it. The presence of this feature, detected in sensitive low frequency observations with the GMRT, therefore, also suggests episodic activity \citep[see also][]{Martini03}. The detection of diffuse low surface brightness emission from old and relic lobes, requires sensitive low frequency observations, as will become available with the SKA1-LOW survey at 120~MHz. 

\section{THE HIGH REDSHIFT UNIVERSE}
It is important to discover high redshift radio galaxies for several reasons. Given the observed correlation between the steepness of the radio spectrum and cosmological redshift (i.e., the $z-\alpha$ correlation), ultra steep spectrum (USS) radio sources are one of the efficient tracers of powerful H$z$RGs \citep{Ishwara-Chandra10,Ker12}. Using deep 150 MHz (1$\sigma\sim$0.7 mJy beam$^{-1}$) observations of LBDS-Lynx field, \citet{Ishwara-Chandra10} reported that among the 150 radio sources with spectra steeper than 1.0, about two-thirds of these are not detected in SDSS, and therefore suggested them to be strong H$z$RGs candidates. In contrast to searches for powerful H$z$RGs from radio surveys of moderate depths, fainter USS samples derived from deeper radio surveys can be useful in finding H$z$RGs at even higher redshifts and in unveiling a population of obscured weaker radio-loud AGN at moderate redshifts. 

Using 325 MHz GMRT observations and 1.4 GHz VLA observations available in two subfields (VLA-VIMOS VLT Deep Survey, VLA-VVDS, and Subaru X-ray Deep Field, SXDF) of the XMM-LSS field, \citet{Singh14} have derived a large sample of 160 faint USS radio sources and characterised their nature. Their study shows that the criterion of ultra steep spectral index remains a reasonably efficient method to select high-$z$ sources even at sub-mJy flux densities. In addition to powerful H$z$RG candidates, their faint USS sample also contains populations of weaker radio-loud AGNs potentially hosted in obscured environments.

Using the $z-\alpha$ correlation, \citet{2011JApA...32..609I,2010MNRAS.405..436I} have found radio galaxies with $z >$ 3 in the deep 150 MHz GMRT-LBDS-Lynx field images with an {\it rms} noise of $\sim$0.7 mJy and an angular resolution of $\sim$17$^{\prime\prime}$. Spectral indices were estimated by cross-correlating the data from these sources with the available data at 327, 610, 1400 and 4860 MHz from radio surveys like the Westerbork Northern Sky Survey and the VLA NVSS and FIRST. They detected about 765 sources in about 15 deg$^2$. 150 of these had spectra steeper than 1.0. Furthermore, about a 100 sources were not detected by SDSS, and are strong candidates for H$z$RGs. As the SKA1 survey will be higher in angular resolution and more sensitive than the GMRT at 150 MHz, we expect to detect a few million candidates for H$z$RGs in SKA level surveys \citep[see Figure~1,][]{2015aska.confE.101J}. 

\subsection{Infrared-Faint Radio Sources}
Recent deep radio surveys combined with auxiliary infrared surveys have discovered the population of Infrared-Faint Radio Sources (IFRS), that are relatively bright radio sources with faint or no counterparts in infrared and optical wavelengths \citep{Middelberg08,Norris11}. IFRS generally exhibit steep radio spectra ($\alpha>1$), high brightness temperatures, T$_{\rm B}$ $\sim10^{6}$~K, and polarisation in the radio, indicating them to be AGN rather than star forming galaxies \citep{Norris07,Banfield11}. 

A search for 3.6~$\mu$m counterparts of 1.4~GHz radio sources in 0.8~deg$^{2}$ of the SXDF,  have yielded seven IFRS distributed over redshifts 1.7 to 4.2, with radio luminosities spanning over 10$^{25}$~W~Hz$^{-1}$ to 10$^{27}$~W~Hz$^{-1}$ (Singh et al. in preparation). This indicates that most, if not all, IFRS are potentially high-redshift ($z>2$) radio-loud AGN suffering from heavy dust extinction. \citet{Zinn11} have compiled a catalogue of 55 IFRS in four deep fields (CDFS (S$_{\rm 1.4\,GHz}\sim186~{\mu}{\rm Jy}$ at 5$\sigma$), ELAIS-S1 (S$_{\rm 1.4\,GHz}\sim160~{\mu}{\rm Jy}$ at 5$\sigma$), FLS (S$_{\rm 1.4\,GHz}\sim105~{\mu}{\rm Jy}$ at 5$\sigma$), and COSMOS (S$_{\rm 1.4\,GHz}\sim65~{\mu}{\rm Jy}$ at 5$\sigma$)), although without redshift estimates. They have shown that the surface number density of IFRS increases with the depth of the radio survey,  and can be best represented as N$_{\rm s}$= ($30.8\pm15.0$) exp{($-0.014\pm0.006$) 5$\sigma$}, where N$_{\rm s}$ is the number of IFRS per deg$^2$, and $\sigma$ is the {\it rms} noise in mJy~beam$^{-1}$. Assuming the SKA1-survey 5$\sigma$ flux density limit of 20~$\mu$Jy~beam$^{-1}$, the best fit equation gives an IFRS surface density of $\sim30.8\pm15$. Therefore, we can expect to detect large number of IFRS using proposed SKA surveys. However, deep optical and IR observations will also be required to estimate their redshifts and to study their dusty host galaxies. Such data should become available from large area surveys like the LSST and VISTA\footnote{Visible and Infrared Survey Telescope for Astronomy}.

\section{DUAL AGN IN GALAXY NUCLEI}
In the merger driven picture of galaxy evolution, as galaxies merge, their supermassive black holes spiral into the centre of the merger remnant forming SMBH pairs \citep[e.g.,][]{Begelman80}. If the black holes are accreting mass, they will form a binary or dual AGN. At  separations of 1 to 10 kpc the accreting SMBH pairs are called dual AGN. At closer separations of a few times 10 parsec or less, the SMBH become gravitationally bound and form binary AGN. At this stage, stars are ejected from the surrounding region until finally the SMBH orbits shrink through the emission of gravitational radiation and the SMBH coalesce \citep{Berczik06}. According to merger scenarios, dual AGN should be common in galaxies. However, the number of confirmed sub-kpc dual AGN is only around 20 \citep{Deane14,Muller15}. The low detection rate is partly because optical or X-ray observations cannot reach the required sub-arsecond resolution, and partly due to dust obscuration. Radio observations are unaffected by dust obscuration and can yield sub-arcsecond images, making them the most efficient frequency for the detection of dual AGN.  

Most of the early dual AGN were detected serendipitously due to radio variability \citep[OJ\,287;][]{Valtonen08}, misaligned radio jets/lobes \citep[e.g., RBS\,797;][]{Gitti13} or double X-ray sources in galaxy nuclei \citep{Fabbiano11}. However, binary/dual AGN can be detected indirectly at optical wavelength using double peaked emission lines in the nuclear spectra of galaxies \citep[e.g.,][]{Liu10}. The double peaks can arise from the Doppler shifts between the emission lines from the two AGN. Large samples of potential dual AGN have been drawn from the SDSS nuclear spectra of galaxies \citep{Ge12} using the [O III] emission line as the principal tracer, since it arises from the NLR in AGN. However, double peaked emission lines can also arise from bipolar outflows in AGN, super-winds from nuclear star formation or emission from gas rotating in accretion disks \citep{Eracleous03,Rosario10,Kharb15b}. 

The SKA1-MID array using band 5 (frequency range $5-14$~GHz) or SKA-VLBI, are needed to detect closely separated binary black holes. These configurations will result in angular resolutions ranging from 0.2 arscec to milli-arcseconds. Through high frequency 8 and 15~GHz observations with the EVLA, Khatun et al. (2016, in preparation) have detected the signatures of a precessing jet, likely arising from a binary black hole system, in a nearby double peaked emission line AGN. Dual or binary AGN at close separations are ideal candidates for gravitational wave studies of merging supermassive black holes. SKA will enable us to detect large samples of such candidates at radio frequencies, for targeted gravitational wave observations.

\section{THE RADIO SKY AT $\mu$Jy LEVELS}
SKA will probe sub-mJy and $\mu$Jy radio source population, which is believed to be largely associated with massive star forming galaxies. However, a significant fraction of the sub-mJy sources are also identified with low-luminosity AGN, with their fraction increasing with flux density \citep{Garrett02}. It is also likely that both starburst and AGN phenomena co-exist in many of the faint systems. The remaining fraction of the faint radio source population are associated with either extremely faint optical identifications, or remain unidentified altogether. 

Deep radio observations of a few fields has led to the detection of many discrete radio sources in a single field of view \citep{Garrett02}. 
Several state-of-the-art modes of the SKA correlator, making full use of the raw data (e.g., mapping out the primary beam response of individual resolution elements in their entirety, or simultaneous multiple-field correlation) coupled with fast data output rates, would be key to achieving high angular resolution images of large areas of the sky in a single pointing. These radio images would then be matched to the large areas of the sky that are routinely being surveyed in great detail by optical and near-IR instruments. {These studies will help us to disentangle the contributions of star-formation and AGN activity to radio emission in galaxies, and lead to the discovery of new types of radio-weak objects. In addition, the `tightness' of the radio - FIR correlation will be tested to higher redshifts, where the correlation is expected to break down due to the dominance of inverse-Compton scattering \citep{Schleicher13}.}  

\section{RADIO CONTINUUM SURVEYS WITH THE GMRT}
The GMRT has been recognised as a pathfinder radio telescope for SKA. Here we briefly describe some of the deep and wide radio continuum surveys that have been carried out with the GMRT {(see Table~\ref{tabgmrt}).}

\subsection{GMRT 325 MHz Survey of Herschel Fields}
The XMM-LSS field has been surveyed with the GMRT at 325~MHz (Wadadekar et al. in preparation). This survey covers an area of $\sim$12 deg$^{2}$ and overlaps fully with the SWIRE and HerMES survey areas in the XMM-LSS field. The 325~MHz GMRT mosaic image has an average {\it rms} noise of $\sim160~\mu$Jy~beam$^{-1}$, while in the central region the {\it rms} noise reaches down to $\sim120~\mu$Jy~beam$^{-1}$, with a synthesized beam of $\sim10\arcsec$.2 $\times$ 7$\arcsec$.9. The 325 MHz survey of XMM-LSS is one of the deepest low-frequency surveys over such a wide sky area and it detects $\sim$2553/3304 radio sources at $\geq$5$\sigma$ with an {\it rms} noise cut-off $\leq$ 200/300 $\mu$Jy~beam$^{-1}$, where a noise image is used for source extraction to account for non-uniformity. It is worth noting that these 325 MHz observations are about five times deeper than the previous 325 MHz observations of the XMM-LSS field \citep[e.g.,][]{Cohen03,Tasse07}, and result in a manifold increase in the source density. 

Nearly 3.8 deg$^{2}$ area of the European Large-Area ISO Survey-North 1 (ELAIS-N1) field has been imaged at 325 MHz with the GMRT \citep{Sirothia09}. It is the most sensitive 325~MHz radio survey with a median {\it rms} noise of $\sim40~\mu$Jy beam$^{-1}$: 1286 sources with a total flux density above $\sim270~\mu$Jy have been detected. 

\subsection{GMRT 610 MHz Survey}
\citet{Garn07} have carried out a 610 MHz survey of the Spitzer extragalactic First Look Survey field (xFLS). This survey covers a total area of $\sim$4 deg$^{2}$ in seven individual pointings with an {\it rms} noise of $\sim30~\mu$Jy~beam$^{-1}$ and resolution of 5$\arcsec$.8 $\times$ 4$\arcsec$.7. This survey has detected a total of 3944 sources above the 5$\sigma$ level. 
{The ELAIS$-$N1, Lockman Hole and VLA$-$VVDS have also been surveyed with the GMRT at 610 MHz (see Table~\ref{tabgmrt})}.  

\subsection{150 MHz GMRT Sky Survey}
The TIFR GMRT Sky Survey (TGSS) is a radio continuum survey at 150~MHz carried out with GMRT. This survey covers $\sim$37,000 deg$^{2}$ of the sky north of declination of $-53$ degrees and reaches an {\it rms} noise level of 5$-$7 mJy beam$^{-1}$, with an angular resolution of $\sim$25$''$ \citep{2016arXiv160304368I}. The first release TGSS has produced a catalog of 0.64 million radio sources at the 7$\sigma$ level. More details about the TGSS can be found at \url{http://tgss.ncra.tifr.res.in/}. This survey and products from it will provide a reference for various new low-frequency telescopes, in particular SKA-LOW. 
{\cite{Intema11} carried out the 150 MHz GMRT survey of the NOVO Bo$\ddot{\rm o}$tes field with an area coverage of $\sim$ 11.3 deg$^{2}$ and an {\it rms} noise of $\sim$1.0 mJy~beam$^{-1}$ .}

\section{NEW DIRECTIONS}
{In addition to expanding the scope of current AGN physics, there are exciting possibilities for new physics that will emerge from SKA. Given the order of magnitude increase in the sensitivity of the upcoming SKA surveys, we are likely to detect new classes of AGN. In particular, the radio-quiet or moderately radio-loud AGN hosted in dusty galaxies at high redshifts. In fact, the modelling of the X-ray background emission predicts a population of obscured AGN \citep{Gilli07}, that has not so far been detected. We can also detect the first generation radio-loud AGN at redshifts $>7$ \citep{Afonso15}. These will place stringent constraints on the relationship between black hole masses, spins, accretion rates, and production of powerful jets, as the supermassive black holes would have formed around then, in the hierarchical galaxy evolution model \citep[e.g.,][]{Li07}.

 Furthermore, radio AGN that are now considered outliers, like hybrid FRI/II sources, X-shaped or highly distorted jet sources, or ``intermediate'' sources that lie close to the radio-loud/radio-quiet divide, or several times re-started AGN sources, could turn out to be norm when observing the radio Universe at high sensitivity. With the detection of many more such sources, it will become possible to create their luminosity functions, and examine their evolution and redshift distributions. Questions of the kind, `are sources at the {\it cusp} transitioning from one AGN class to another ?', will be answered. For instance, it has been hypothesised that FRII radio galaxies evolve into FRI radio galaxies \citep[e.g.,][]{Baum95}. Are hybrid FRI/II sources then the transitioning objects ? This can be rigorously tested when more hybrid and FRI sources with weak diffuse lobes are discovered at higher redshifts. These discoveries are expected to fundamentally alter our view of AGN, of galaxies, and of the Universe.}

\section{CONCLUDING REMARKS}
We have summarised various science interests of a large fraction of the AGN community in India. Whether it comes to detecting ``intermediate'' sources between the radio-loud and radio-quiet classes, or ``intermediate'' sources between FRIs and FRIIs, relic emission from previous activities of the AGN, giant radio jets in spiral galaxies, double-double or triple-double radio galaxies, or faint ultra-steep spectrum sources at high redshifts, higher sensitivity data that are at present unavailable, are required. The various SKA configurations and the upcoming SKA surveys will meet these needs. VLBI is required to probe the regions close to the central black holes. Again, an increase in sensitivity is crucial. SKA-VLBI which could include large single antennas like the upcoming FAST radio telescope, will be able to study parsec and sub-parsec-scale radio jets and their magnetic field structures in low luminosity AGN, which have so far remained below the sensitivity limits of the current VLBI arrays. In short, several burning questions on AGN physics, the answers to which have primarily been limited by the lack of resolution, sensitivity or statistically significant number of sources, will be addressed directly by SKA. Many more new leading questions on AGN physics will hopefully be raised.

\acknowledgments
We thank the anonymous referee for their careful reading of our manuscript and providing insightful suggestions that have improved this paper. MP is thankful to the CEFIPRA organization for the  Franco-Indian research grant. 


\bibliographystyle{apj}
\bibliography{ms}

\begin{table}
\small
\centering
\caption{GMRT radio continuum surveys of extragalactic fields}
\begin{tabular}{llllll}
\hline\hline
{Frequency}   &    {Field}      &      {Area}     &     {{\it rms} noise}   &    beam-size    &    Reference     \\ 
{(MHz)}       &      {}         &   {(deg$^{2}$)} & {(mJy beam$^{-1}$)} &                 &                  \\ \hline
610           & Lockman Hole    &   0.2           &    0.015          & 7$\arcsec$.1 $\times$ 6$\arcsec$.5  &  \cite{Ibar09} \\
610           &  VLA-VVDS       &   1.0           &    0.05           & 6$\arcsec$.0 $\times$ 5$\arcsec$.0  &  \cite{Bondi07}  \\
610           &   xFLS          &   4.0           &    0.03           & 5$\arcsec$.8 $\times$ 4$\arcsec$.7  &  \cite{Garn07}   \\
610           & Lockman Hole    &   8.0           &    0.08           & 6$\arcsec$.0 $\times$ 5$\arcsec$.0  &  \cite{Garn10}   \\
610           & ELAIS-N1        &   9.0           &  0.04 $-$ 0.07    & 6$\arcsec$.0 $\times$ 5$\arcsec$.0  &  \cite{Garn08}    \\
610           & ELAIS-N1        &   1.0           &    0.01           & 6$\arcsec$.0 $\times$ 5$\arcsec$.0  &  \cite{Taylor16}    \\
325           & ELAIS-N1        &   1.1           &    0.04           & 9$\arcsec$.4 $\times$ 7$\arcsec$.4  &  \cite{Sirothia09} \\
325           & XMM-LSS         &   12            &    0.15           & 10$\arcsec$.2 $\times$ 7$\arcsec$.9 & Wadadekar et al. in preparation  \\
325           & H-ATLAS         &   90            &    1.0            &  14$\arcsec$ $-$ 24$\arcsec$        &  \cite{Mauch13}   \\
150  & NOVO Bo$\ddot{\rm o}$tes &   11.3          &    1.0            &  26$\arcsec$ $\times$ 22$\arcsec$   &  \cite{Intema11}  \\ 
150           & TGSS            &   36900         &    3.5            &  25$\arcsec$ $\times$ 25$\arcsec$   &  \cite{Intema16}  \\ \hline
\end{tabular}
\vspace{0.35cm}
\label{tabgmrt}
\end{table}

\end{document}